\documentclass[a4paper, 10pt, conference]{ieeeconf}      

\IEEEoverridecommandlockouts                              
\usepackage{amsfonts}
\usepackage{graphicx}
\usepackage{amsmath}

\usepackage{ifthen}
\usepackage{color}
\usepackage{cite}

\def\eq#1{\begin{equation}#1\end{equation}}
\newcommand{\R}{{\rm I\!R}}

\def\rep#1{(\ref{#1})}
\def\scr#1{{\cal #1}}
\newtheorem{theorem}{Theorem}
\newtheorem{lemma}{Lemma}
\newtheorem{proposition}{Proposition}

\newtheorem{assumption}{Assumption}
\newtheorem{remark}{Remark}
\newtheorem{example}{Example}
\def\qed{ \rule{.08in}{.08in}}

\newcommand{\bbb}{\mathbb}

\newcommand{\1}{\mathbf{1}}
\newcommand{\0}{\mathbf{0}}

\title{\LARGE \bf Discrete-Time Polar Opinion Dynamics with Susceptibility
}

\author{Ji Liu, Mengbin Ye, Brian D.O. Anderson, Tamer Ba\c{s}ar, Angelia Nedi\'c
\thanks{This research was supported in part by ONR MURI Grant N00014-16-1-2710, and in part by the Australian Research Council under grants
DP-130103610 and DP-160104500, and Data61-CSIRO.
J. Liu is with the Department of Electrical and Computer Engineering at Stony Brook University
(\texttt{ji.liu@stonybrook.edu}).
M. Ye is with the Research School of Engineering at Australian National University, and supported by an Australian Government
Research Training Program (RTP) Scholarship (\texttt{mengbin.ye@anu.edu.au}).
B.D.O. Anderson is with the Research
School of Engineering, Australian National University 
(\texttt{Brian.Anderson@anu.edu.au}), Hangzhou Dianzi University, Hangzhou, China, and Data61-CSIRO (formerly NICTA Ltd.) in Canberra, A.C.T., Australia.
T. Ba\c{s}ar is with the Coordinated Science Laboratory at
University of Illinois at Urbana-Champaign (\texttt{basar1@illinois.edu}).
A. Nedi\'c is with the School of Electrical,
Computer and Energy Engineering at Arizona State University (\texttt{angelia.nedich@asu.edu}).
}
}

\begin{document}

\maketitle
\thispagestyle{empty}
\pagestyle{empty}

\begin{abstract}

This paper considers a discrete-time opinion dynamics model in which each individual's susceptibility to being influenced by others is dependent on her current opinion. We assume that the social network has time-varying topology and that the opinions are scalars on a continuous interval. We first propose a general opinion dynamics model based on the DeGroot model, with a general function to describe the functional dependence of each individual's susceptibility on her own opinion, and show that this general model is analogous to the Friedkin-Johnsen model, which assumes a constant susceptibility for each individual. 
We then consider two specific functions in which the individual's susceptibility depends on the \emph{polarity} of her opinion, and provide motivating social examples. First, we consider stubborn positives, who have reduced susceptibility if their opinions are at one end of the interval and increased susceptibility if their opinions are at the opposite end. A court jury is used as a motivating example. Second, we consider stubborn neutrals, who have reduced susceptibility when their opinions are in the middle of the spectrum, and our motivating examples are social networks discussing established social norms or institutionalized behavior. For each specific susceptibility model, we establish the initial and graph topology conditions in which consensus is reached, and develop necessary and sufficient conditions on the initial conditions for the final consensus value to be at either extreme of the opinion interval. Simulations are provided to show the effects of the susceptibility function when compared to the DeGroot model.   

\end{abstract}


\section{Introduction}
The problem of opinion dynamics, which considers how an individual's opinion forms and evolves through interactions with others in a social network, has been widely studied in the social sciences for decades. The classical discrete-time DeGroot model, in which each individual updates her opinion by taking a convex combination of the opinions of her neighbors at each time step, is perhaps one of the most well known models \cite{degroot}. This model is closely related to discrete-time linear consensus algorithms, which have been heavily studied in multi-agent coordination literature \cite{Ts3,vicsekmodel,reza1,ReBe05,luc,blondell,survey,touri2014,cdc14,rate}. Since the time the DeGroot model was proposed, numerous other models have been introduced, in both continuous- and discrete-time setting. These various models, which describe the opinion formation process in the context of different social cognitive processes, all attempt to understand the formation and evolution of opinions in social networks of all sizes, and explain observed social phenomena such as polarization or attitude extremity \cite{osgood,miller,pomerantz,bassili}, and subculture formation \cite{mas2014cultural,duggins2017_psych_opdyn}.


There are many variants of the DeGroot model for opinion dynamics. 
The Altafini model, which suggests the interactions between individuals can be cooperative or antagonistic, has been studied as a discrete-time process in \cite{weiguo,modulus,lift,cdc15},
and the continuous-time counterpart has been considered in \cite{altafini,signcdc13,mingtac,acc16}. It is notable because the model links the limiting opinion behavior with the structural balance of the graph representing the social network. Some other models primarily focus on linking the limiting opinion behavior with a social process. For example, the Hegselmann-Krause model shows the social cognitive process of homophily is linked to fact that opinions in the social network eventually form clusters \cite{krause1,krause,etesami}. It was shown in \cite{mas2014cultural} that an individual's desire to strive for uniqueness generated persistent subcultures which formed and vanished over time. On the other hand, an individual conforming to a social norm generated pluralistic ignorance \cite{duggins2017_psych_opdyn}. Finally, some models attempt to link final opinion behavior to a combination of social processes and the underlying network structure. The Friedkin-Johnsen model \cite{magazine,friedkin2011social_book} considered individual susceptibility to influence and shows that opinions reach a persistent diversity under general graph structures. The DeGroot-Friedkin model \cite{jia} studied an individual's ability to reflect on her impact in the opinion formation process, and showed her social power depended on the graph structure.

A key aspect of the DeGroot model is the interpersonal influence, which describes the amount of influence each individual's neighbors have in determining that individual's new opinion. Some of the results consider arbitrary, time-varying interpersonal influence, e.g. \cite{luc,modulus,mingtac}. However, many of the aforementioned models consider influence determined by a social process, e.g. homophily \cite{etesami}, social distancing \cite{mas2014cultural,duggins2017_psych_opdyn}, conformity \cite{duggins2017_psych_opdyn}, desire for uniqueness \cite{mas2014cultural}, biased assimilation \cite{dandekar2013biased_degroot}, or reflected self-appraisal \cite{jia}. Because the social process is often dependent on the states, i.e. opinions (which change with time), then necessarily the interpersonal influences are state-dependent, and thus time-varying. In a recent paper \cite{polar}, a continuous-time model has been proposed for fixed social network topology which considers, separately, three different cognitive processes to drive the influence change. In \cite{polar}, the term ``polar opinion dynamics'' relates to the fact that the level of influence is dependent on how extreme, i.e. polar, an individual's opinion is. 


In this paper, we study a discrete-time opinion dynamics model where an individual's susceptibility to influence is dependent on her current opinion, and allow the social network topology to vary over time. We first propose a general model, show it can be considered as a generalization of both the DeGroot model and the Friedkin-Johnsen model \cite{magazine,friedkin2011social_book}, and establish some general properties of the model. We then investigate discrete-time versions of two of the three cognitive process introduced in \cite{polar}, bearing in mind that discrete-time models are more appropriate to describe opinion dynamics, at least from the point that individuals change their minds from time to time, instead of continuously. In addition, we provide social examples from existing literature to motivate these susceptibility functions. For each function, we provide sufficient conditions on the graph topology and initial opinions for the social network to reach a consensus. Importantly, we establish necessary and sufficient conditions for the social network to hold opinions at either extremes of the opinion interval, 
whereas \cite{polar} provided only sufficient conditions, or no conditions at all for extremity of the final consensus in the continuous-time model. Lastly, it turns out that while some of the limiting behaviors are similar to the continuous-time model, in other cases, the limiting behaviors are not the same.

The remainder of this paper is organized as follows.
Some notations and preliminaries are introduced in Section~\ref{notion}.
In Section~\ref{sec:model}, the discrete-time polar opinion dynamics model with susceptibility is introduced.
The main results of the paper are presented in Section~\ref{sus},
which are illustrated and compared with the DeGroot model via simulations in Section~\ref{sim}.
The paper ends with some concluding remarks in Section~\ref{ending}.

\subsection{Preliminaries} \label{notion}

For any positive integer $n$, we use $[n]$ to denote the index set $\{1,2,\ldots,n\}$.
We view vectors as column vectors and write $x^\top$ to denote the transpose of a vector $x$.
For a vector $x$, we use $x_i$ to denote the $i$th entry of $x$.
For any matrix $M\in\bbb R^{n\times n}$, we use $m_{ij}$ to denote its $ij$th entry.
A nonnegative $n\times n$ matrix is called a stochastic matrix if its row sums are all equal to $1$.
We use $\0$ and $\1$ to denote the
vectors whose entries all equal 0 and 1, respectively, and $I$
to denote the identity matrix, while the dimensions of the
vectors and matrices are to be understood from the context.
For any real number $x$, we use $|x|$ to denote the absolute value of $x$.
For any two real vectors $a,b\in\R^n$, we write $a\geq b$ if
$a_{i}\geq b_{i}$ for all $i\in[n]$,
$a>b$ if $a\geq b$ and $a\neq b$, and $a \gg b$ if $a_{i}> b_{i}$ for all $i\in[n]$.
For any two sets $\scr A$ and $\scr B$, we use $\scr A\setminus \scr B$ to denote the set of elements
in $\scr A$ but not in $\scr B$.
The graph of an $n\times n$ matrix  $M$ with real-valued entries is an $n$-vertex directed
 graph 
 defined so that $(i,j)$ is an arc from vertex $i$ to vertex $j$  in the graph whenever
  the $ji$th entry of
 $M$ is nonzero.
We will use the terms ``individual'' and ``agent'' interchangeably.


\section{The General Model} \label{sec:model}
In this section, we propose a general model for describing opinion dynamics where each individual's susceptibility to being influenced by others is affected by some social process, and give some results on the trajectories of the opinions. In the next section, we shall propose two specific models to describe two different variants of a social process.

Consider a social network of $n>1$ agents, labeled $1$ through $n$, discussing opinions on a given topic.\footnote{
The purpose of labeling of the agents is only for convenience. We do not require a global labeling of the agents
in the network. We only assume that each agent can identify her own neighbors.}
Each agent $i$ can only learn, and be influenced by, the opinions of certain other agents called the
neighbors of agent $i$. Neighbor relationships among the $n$ agents are
described by a directed graph $\bbb N(t)$, called the neighbor graph, which may change over time.
Agent $j$ is a neighbor of agent $i$ at time $t$ whenever
$(j,i)$ is an arc in $\bbb N(t)$. Thus, the directions of arcs indicate
the directions of information flow (specifically opinion flow).
For convenience, we assume that each agent is always a neighbor of herself.
Thus, $\bbb N(t)$ has self-arcs at all $n$ vertices for all time $t$.
Each agent $i$ has control over a real-valued quantity $x_i$, called
agent $i$'s opinion.

In the time-varying DeGroot model\footnote{The original DeGroot model was proposed for a fixed graph \cite{degroot}. Subsequent results expanded this to time-varying graphs, e.g. \cite{vicsekmodel,luc}.},
each agent $i$ updates her opinion at each discrete time $t\in\{1,2,\ldots\}$ by setting
\eq{x_i(t+1) = \sum_{j\in\scr{N}_i(t)} w_{ij}(t)x_j(t), \;\;\;\;\; i\in [n],\label{degroot}}
where $\scr{N}_i(t)$ denotes the set of neighbors of agent $i$ at time $t$ including $i$ herself,
and $w_{ij}(t)$ are positive influence weights
satisfying $\sum_{j\in\scr{N}_i(t)} w_{ij}(t)=1$ for all $i\in[n]$ and time $t$. We assume that the weights $w_{ij}(t)$ change in a manner which is entirely independent of $x_k(t), \forall\,k \in [n]$.
We rewrite the above model as
\begin{eqnarray*}
x_i(t+1) &=& x_i(t) + \left(\sum_{j\in\scr{N}_i(t)} w_{ij}(t)x_j(t) - x_i(t)\right) \\
 &=& x_i(t) + \sum_{j\in\scr{N}_i(t)} w_{ij}(t)\left(x_j(t)-x_i(t)\right),
\end{eqnarray*}
with the second equality obtained by noting that $\sum_{j\in\scr{N}_i(t)} w_{ij}(t)=1$, and define
$$u_i(t)=\sum_{j\in\scr{N}_i(t)} w_{ij}(t)\left(x_j(t)-x_i(t)\right).$$
Then, $u_i(t)$ represents the influence of agent $i$'s neighbors, which generates a change in the opinion of agent $i$, i.e. $u_i(t)$ can be viewed as the control input of agent $i$ at time $t$.
We now suppose that agent $i$ may not fully accept the influence of her neighbors, and her openness to influence, or susceptibility, is captured by the real-valued function $f_i(x_i(t))$. We make the following assumption on $f_i(x_i(t))$:
\begin{assumption}
The susceptibility function $f_i(x_i(t))$ takes on values in $[0,1]$.
\end{assumption} 
We consider the following model for opinion dynamics
with susceptibility:
\eq{
x_i(t+1) = x_i(t) + f_i(x_i(t))\sum_{j\in\scr{N}_i(t)} w_{ij}(t)\left(x_j(t)-x_i(t)\right).
\label{model}}
If at time $t$ $f_i(x_i(t))=1$, agent $i$ fully accepts
her neighbors' influence at time $t$, and the model reduces to the DeGroot model.
In the case when $f_i(x_i(t))=0$, agent $i$ will ignore her neighbors and
not change her opinion at time $t$; in such a case, the agent is sometimes called stubborn \cite{srikant,magazine}.
It is worth emphasizing that an agent's susceptibility function
depends on its current opinion, i.e. state. This is consistent with the many works discussed in the introduction, which considers social cognitive processes which are dependent on the individual's opinion and, in some instances, the opinions of her neighbor.

There is also another interpretation of the model.
Inspired by the Friedkin-Johnsen model \cite{magazine,friedkin2011social_book,johnsen},
we assume that each agent $i$ updates her opinion as a convex combination
of her own opinion and the weighted average of her neighbors' opinions;\footnote{
In the Friedkin-Johnsen model, $x_i(t)$ of the second term on the right of \rep{ben} is replaced by $x_i(0)$.} specifically,
\begin{equation}
x_i(t+1) = \lambda_i \sum_{j\in\mathcal{N}_i(t)} w_{ij}(t) x_j(t) + (1-\lambda_i) x_i(t),
\label{ben}\end{equation}
where the constant $\lambda_i \in [0,1]$ is agent $i$'s susceptibility or openness to being influenced by her neighbors' opinions. Let us replace $\lambda_i$ with the state-dependent susceptibility function $f_i(x_i(t))$. It follows that
\begin{align}
& x_i(t+1) \nonumber\\
&  = f_i(x_i(t))\sum_{j\in\mathcal{N}_i(t)} w_{ij}(t) x_j(t) + (1-f_i(x_i(t))) x_i(t) \label{convex}\\
& = x_i(t) - f_i(x_i(t)) \left( x_i(t) - \sum_{j\in \mathcal{N}_i(t)} w_{ij}(t) x_j(t) \right)  \nonumber\\
& = x_i(t) - f_i(x_i(t)) \left( \sum_{j\in \mathcal{N}_i(t)} w_{ij}(t) \left( x_i(t) - x_j(t) \right) \right) \nonumber\\
& = x_i(t) + f_i(x_i(t)) u_i(t), \nonumber
\end{align}
which is the same as \rep{model}.

\begin{remark}
We note here that there are two types of time-dependency in the influence term $u_i(t)$. Firstly, we have assumed that the influence weights $w_{ij}(t)$ may be time-varying \emph{but state-independent.} This may occur in situations where an individual decides to at time $t$, stop sharing her opinion, stop listening to certain neighbors, start listening to other neighbors, or adjust weight magnitudes (perhaps by becoming more persuasive) etc. This differs from the continuous-time work in \cite{polar}, which considered static influence weights. The second is time-dependency arising from the fact that the susceptibility of individual $i$, $f_i(x_i(t))$, is state-dependent. In the Friedkin-Johnsen model, susceptibility was assumed to be constant. While some papers have studied state-dependent susceptibility in discrete-time models, they provide only simulations, and have not provided rigorous analysis or considered time-varying influence weights $w_{ij}(t) $\cite{duggins2017_psych_opdyn}.
\hfill$\Box$
\end{remark}

In this paper, we assume that all the initial opinions $x_i(0)$, $i\in[n]$, lie in the interval $[-1,1]$,
where $-1$ and $1$ represent the extreme positive and negative opinions, respectively. Such a scaling is typical in opinion dynamics problems where $x_i$ may represent individual $i$'s attitude towards an idea, e.g. the legalization of recreational marijuana, with $x_i = 1$ maximally supportive and $x_i = -1$ maximally opposing.
The following lemma shows that $[-1,1]$ is an invariant set of each agent's opinion dynamics
given by \rep{model}.

\begin{lemma}
Suppose that each agent $i$ follows the update rule \rep{model}
and that $x_i(0)\in[-1,1]$ for all $i\in[n]$. Then,
$x_i(t)\in[-1,1]$ for all $i\in[n]$ and time $t$.
\label{invariant}
\end{lemma}

{\em Proof:}
From \rep{convex} and the assumption that $f_i(x_i(t))\in[0,1]$,
each agent $i$'s updated value $x_i(t+1)$ is a convex combination of
$x_i(t)$ and  $\sum_{j\in\mathcal{N}_i(t)} w_{ij}(t) x_j(t)$; i.e. a convex combination
of the current opinions of her neighbors.
Using induction, it is easy to see that if $x_i(0)\in[-1,1]$ for all $i\in[n]$,
it follows that $x_i(t)\in[-1,1]$ for all $i\in[n]$ and time $t$.
\hfill$\qed$

This shows that individual $i$'s opinion will remain bounded from above and below, and ensures that an individual's opinion cannot become increasingly extreme in either the positive or negative direction.
More can be said about the most extreme opinions in the social network. Specifically, the most negative and positive opinions will never become more negative and more positive, respectively.

\begin{lemma}
Suppose that each agent $i$ follows the update rule \rep{model}.
Then, $x_{{\rm min}}(t) = \min_i x_i(t)$ is nondecreasing
and $x_{{\rm max}}(t) = \max_i x_i(t)$ is nonincreasing as $t$ increases.
\label{mono}\end{lemma}

{\em Proof:}
From \rep{model}, it follows that
\begin{eqnarray*}
&&x_i(t+1) \\
&=& x_i(t) + f_i(x_i(t))\left(\sum_{j\in\scr{N}_i(t)} w_{ij}(t)x_j(t)-x_i(t)\right) \\
&=& (1-f_i(x_i(t)))x_i(t) + f_i(x_i(t))\sum_{j\in\scr{N}_i(t)} w_{ij}(t)x_j(t)
\end{eqnarray*}
The above set of $n$ equations can be combined into state form. Toward this end,
let $x(t)$ be the vector in $\R^n$ whose $i$th entry equals $x_i(t)$,
$F(x(t))$ be the $n\times n$ diagonal matrix whose $i$th diagonal entry equals $f_i(x_i(t))$ with $0 \leq f_i \leq 1 $,
and $W(t)$ be the $n\times n$ matrix whose $ij$th entry equals $w_{ij}(t)$.
Then, it follows that
\begin{eqnarray}
x(t+1) &=& (I-F(x(t)))x(t)+F(x(t))W(t)x(t) \nonumber\\
&=& S(x(t),t)x(t), \label{nonlinear}
\end{eqnarray}
where
$$S(x(t),t) =  I-F(x(t))+F(x(t))W(t).$$
It is worth noting that $S(x(t),t)$ is a function of $x(t)$ as $F(x(t))$ is so.
Thus, \rep{nonlinear} is a nonlinear system.
From Lemma~\ref{invariant} and Assumption~\ref{weight},
$S(x(t),t)$ is a nonnegative matrix for all time $t$.
Since $W(t)$ is a stochastic matrix, it follows that
\begin{eqnarray*}
S(x(t),t)\1 &=& \left(I-F(x(t))+F(x(t))W(t)\right)\1 \\
&=& \1 - F(x(t))\1 + F(x(t))\1 \\
&=& \1,
\end{eqnarray*}
which implies that $S(x(t),t)$ is a stochastic matrix for all time $t$.
Thus, each $x_i(t+1)$ is a convex combination of all $x_i(t)$, $i\in[n]$,
which implies that $x_{{\rm min}}(t) = \min_i x_i(t)$ is nondecreasing
and $x_{{\rm max}}(t) = \max_i x_i(t)$ is nonincreasing as $t$ increases.
\hfill$\qed$

We also impose the following set of
assumptions on the weights $w_{ij}(t)$ throughout the rest of the paper.

\begin{assumption}
For all $i\in[n]$ and $t$, there hold $w_{ij}(t)>0$
if $j\in\scr N_i(t)$ and $w_{ij}(t)=0$ otherwise.
There exists a positive number $\beta$ such that, for all $i\in[n]$ and $j\in\scr N_i(t)$, if $w_{ij} \neq 0$, then $w_{ij} \geq \beta$.
For all $i\in[n]$ and $t$, there holds
$\sum_{j\in\scr{N}_i(t)} w_{ij}(t) =1$.
\label{weight}
\end{assumption}

Such a set of assumptions
implies that $W(t)=[w_{ij}(t)]$ is a stochastic matrix for all time $t$,
and was widely used in the DeGroot (and consensus) studies
\cite{rate}. For the (time-varying) DeGroot model \rep{degroot},
we have a standard result, which we state after first defining some connectivity conditions for time-varying graphs.

A directed graph $\mathbb{G}$ is {\em strongly connected} if there is a directed path between
each pair of its distinct vertices.
We say that a finite sequence of directed graphs $\bbb{G}_1,\bbb{G}_2,\ldots,\bbb{G}_m$ with the same vertex set
is {\em jointly strongly connected} if the union\footnote{
The union of a finite sequence of directed graphs with the same vertex set
is a directed graph with the same vertex set and the arc set which is the union
of the arc sets of all directed graphs in the sequence.
}
of the directed graphs in this sequence is strongly connected.
We say that an infinite  sequence of directed graphs $\mathbb{G}_1, \mathbb{G}_2,\ldots$ with the same vertex set
is {\em repeatedly jointly strongly connected}
if there exist positive integers $p$ and $q$ for which
each finite sequence  $\mathbb{G}_{q+kp},\mathbb{G}_{q+kp+1},\ldots,\mathbb{G}_{q+(k+1)p-1}$,
$k\geq 0$, is jointly strongly connected.
Repeatedly jointly strongly connected graphs are equivalent to
so-called ``$B$-connected''
graphs in the consensus literature \cite{nedic3} whose definition is in a slightly different form.

\begin{proposition}
{\rm (Theorem 2 in \cite{rate})}
Suppose that Assumption \ref{weight} holds.
If the sequence of neighbor graphs $\bbb N(1),\bbb N(2),\ldots$ is
repeatedly jointly strongly connected\footnote{
The result still holds if $\bbb N(1),\bbb N(2),\ldots$ is
repeatedly jointly rooted \cite{cdc14}.}, then all $x_i(t)$, $i\in[n]$, in \rep{degroot}
will reach a consensus exponentially fast as $t\rightarrow\infty$ for all initial conditions.
\label{consensus}\end{proposition}
It is worth noting that for each time $t$, neighbor graph $\bbb N(t)$ is the same
as the graph of weight matrix $W(t)=[w_{ij}(t)]$.

We will use this result in our analysis of the model \rep{model}.

\section{Susceptibility Functions} \label{sus}

In this section, we will consider two specific susceptibility functions,
and study the behavior of the corresponding models. We give motivating examples from sociology for the susceptibility functions taking on these specific forms.

\subsection{Stubborn Positives}

We begin with the case where $f_i(x_i(t)) = \frac{1}{2}(1-x_i(t))$, for all $i\in [1]$. An agent which has this susceptibility function is called a stubborn positive.

\subsubsection*{Motivation}
Note here that for stubborn positive agents, her susceptibility decreases as $x_i \to 1$, and increases as $x_i \to -1$. In other words, the closer the agent is to a ``positive opinion'' (respectively a ``negative opinion''), the more stubborn or unwilling (respectively more open or susceptible) she is to changing her opinion. Our motivating example is a jury panel. The paper \cite{waters2009jury} conducted extensive surveys of criminal juries after trials were complete. A clear pattern was observed: a juror was more likely to be extremely stubborn when believing the defendant should be acquitted, than when believing the defendant should be convicted. In our context, a juror with $x_i = 1$ is maximally supportive of acquitting the defendant, while a juror with $x_i = -1$ is maximally opposing acquittal (and thus supportive of convicting)\footnote{We note here that this is the opinion of the juror, as opposed to the final action taken by the juror. An individual may privately take one opinion and express another due to the social circumstances \cite{asch1951group_pressure_effects}.}. It was suggested that this asymmetric stubbornness arose from the fact that a false conviction carried an enormous amount of consequence for defendants in criminal cases, e.g. a prison sentence. In summary, scenarios which involve social networks with stubborn positives can arise in discussions where the outcome for one result has drastically different severity of consequences compared to the opposite result.

\subsubsection*{Analysis}
In this case, from \rep{model}, each agent $i$ updates her opinion by setting
\begin{eqnarray}
&&x_i(t+1) \nonumber\\
&=& x_i(t) + \frac{1}{2}(1-x_i(t))\sum_{j\in\scr{N}_i(t)} w_{ij}(t)\left(x_j(t)-x_i(t)\right) \nonumber \\
&=& x_i(t) + \frac{1}{2}(1-x_i(t))\left(\sum_{j\in\scr{N}_i(t)} w_{ij}(t)x_j(t)-x_i(t)\right) \nonumber \\
&=& \frac{1}{2}(1+x_i(t))x_i(t)+\frac{1}{2}(1-x_i(t))\sum_{j\in\scr{N}_i(t)} w_{ij}(t)x_j(t). \nonumber \\
&&
\label{positive}
\end{eqnarray}
It follows that
\begin{eqnarray}
x(t+1) &=& \frac{1}{2}(I+X(t))x(t) + \frac{1}{2}(I-X(t))W(t)x(t) \nonumber\\
&=& \left(\frac{1}{2}(I+X(t)) + \frac{1}{2}(I-X(t))W(t)\right) x(t), \nonumber \\
&& \label{positive1}
\end{eqnarray}
where $X(t)$ is the $n\times n$ diagonal matrix whose $i$th diagonal entry equals $x_i(t)$.


The following theorem characterizes the limiting behavior of system \rep{positive1}.

\begin{theorem}
Suppose that Assumption \ref{weight} holds and
that the sequence of neighbor graphs $\bbb N(1),\bbb N(2),\ldots$ is
repeatedly jointly strongly connected.
If $x_i(0) < 1$ for all $i\in[n]$, then all $x_i(t)$, $i\in[n]$, in \rep{positive}
will reach a consensus exponentially fast at some value in the interval $[-1,1)$;
moreover, in this case, the consensus value equals $-1$ if, and only if, $x_i(0) = -1$ for all $i\in[n]$.
If $x_i(0) = 1$ for at least one $i\in[n]$, then all $x_i(t)$, $i\in[n]$, in \rep{positive}
will reach a consensus at value $1$.
\label{mainpositive}\end{theorem}

{\em Proof:}
From \rep{positive1}, $x(t+1)=S(x(t),t)x(t)$ where
\eq{S(x(t),t) =  \frac{1}{2}(I+X(t)) + \frac{1}{2}(I-X(t))W(t).\label{xxx}}
From the proof of Lemma \ref{mono}, $S(x(t),t)$ is a stochastic matrix for all time $t$.

First suppose that $x_i(0) < 1$ for all $i\in[n]$.
There must exist a positive number $\alpha$ such that $x_i(0)\le 1-\alpha$ for all $i\in[n]$.
It follows from Lemma \ref{mono} that $x_i(t)\le 1-\alpha$ for all $i\in[n]$ and time $t$.
From \rep{xxx}, we obtain the following two inequalities for $s_{ij}(t)$, the entries of $S(x(t),t)$.
For each diagonal entry,
\begin{eqnarray*}
s_{ii}(t) &=& \frac{1}{2}(1+x_i(t)) + \frac{1}{2}(1-x_i(t))w_{ii}(t) \\
&\ge& \frac{1}{2}(1-x_i(t))w_{ii}(t) \\
&\ge & \frac{1}{2}\alpha\beta,
\end{eqnarray*}
where the last inequality makes use of Assumption~\ref{weight} and the fact that we assumed every node in $\bbb N(t)$ has a self-loop, for all $t$.
For each off-diagonal entry,
$$s_{ij}(t)=\frac{1}{2}(1-x_i(t))w_{ij}(t).$$
Thus, $s_{ij}(t)$ is nonzero if and only if $w_{ij}(t)$ is nonzero (because $1-x_i(t) \geq \alpha > 0$),
which implies that the graph of $S(x(t),t)$ has the same edge and vertex set (but with different edge weights) as the graph of $W(t)$, as well as neighbor graph $\bbb N(t)$.
Moreover, it can be seen that when $s_{ij}(t)>0$, it must hold that $s_{ij}(t)\ge \frac{1}{2}\alpha \beta$.
From Proposition \ref{consensus}, all $x_i(t)$, $i\in[n]$, will reach a consensus exponentially fast.
Since $-1\le x_i(t)\le 1-\alpha$ for all $i\in[n]$ and time $t$, the consensus value must lie in $[-1,1)$.

Next we show that all $x_i(t)$, $i\in[n]$, in \rep{positive}
will reach a consensus at value $-1$ if and only if $x_i(0) = -1$ for all $i\in[n]$.
Suppose that, to the contrary, there exists at least one $i\in[n]$ such that $x_i(0)>-1$.
From \rep{positive}, since $x_j(t)\ge -1$ for all $j\in\scr N_i(t)$,
there holds
\begin{eqnarray*}
x_i(t+1) &\ge&  \frac{1}{2}(1+x_i(t))x_i(t) - \frac{1}{2}(1-x_i(t)).
\end{eqnarray*}
Suppose that $x_i(t)>-1$. Then, it follows that
$$x_i(t+1) > -\frac{1}{2}(1+x_i(t)) - \frac{1}{2}(1-x_i(t)) = -1,$$
which implies that
if $x_i(0)>-1$, then $x_i(t)>-1$ for all $t$.
Let $\scr S(t)$ denote the set of agents whose opinions are greater than $-1$ at time $t$.
From the hypothesis and preceding discussion, $\scr S(t)$ is nonempty for all time $t$.
Let $\bar{\scr  S}(t)=[n]\setminus \scr S(t)$ be the set of agents whose opinions equal $-1$ at time $t$.
If $\bar{\scr  S}(t)$ is empty, i.e.,
$x_i(0) > -1$ for all $i\in[n]$, then from Lemma \ref{mono}, the system cannot reach a consensus at $-1$.
Suppose that $\bar{\scr  S}(t)$ is nonempty, i.e., there exists at least one agent whose initial opinion is $-1$.
Since the sequence of neighbor graphs is repeatedly jointly strongly connected,
there must exist a finite time $\tau$ and an agent $j\in\bar{\scr S}(\tau)$  such that it has a neighbor
$k\in\scr S(\tau)$, i.e., $x_j(\tau)=-1$ and $x_k(\tau)>-1$ with $k\in\scr N_j(\tau)$.
From \rep{positive}, it can be seen that $x_j(\tau+1)>-1$.
Using the same arguments, there exists a finite time $\bar \tau$ such that
$x_i(\bar\tau)>-1$ for all $i\in[n]$, which contradicts the hypothesis that
all $x_i(t)$, $i\in[n]$, will reach a consensus at $-1$. Therefore,
the consensus at $-1$ will be reached if and only if $x_i(0) = -1$ for all $i\in[n]$.

Now we consider the case when $x_i(0) = 1$ for at least one $i\in[n]$.
Consider the
Lyapunov function
$$V(x(t)) = 1-\min_{i\in[n]}x_i(t).$$
From \rep{positive}, if $x_i(t)=1$, then $x_i(t+1)=1$, which implies that
if $x_i(0) = 1$, then $x_i(t)=1$ for all time $t$.
Thus, if $x_i(t)=1$ for all $i\in[n]$ at some time $t$, then
$V(x(t))=0$ and $V(x(t+1))=0$.
Suppose that 
there exists at least one agent $i$ such that $x_i(t)<1$ at a specific time $t$.
In this case,  $\min_{i\in[n]}x_i(t)<1$ and thus $V(x(t))>0$.
From Lemma~\ref{mono}, there holds $x_j(\tau)\ge \min_{i\in[n]}x_i(t)$ for all $j\in[n]$ and $\tau\ge t$.
Let $\scr M(t)$ be the set of agents whose opinions are the smallest at time $t$,
i.e., $x_j(t)=\min_{i\in[n]}x_i(t)$ for each $j\in\scr M(t)$.
Since $x_i(0) = 1$ for at least one $i\in[n]$,
it follows that $[n]\setminus\scr M(t)$ is nonempty for all $t$.
Since the sequence of neighbor graphs is repeatedly jointly strongly connected,
there must exist a finite time $\tau\ge t$ and an agent $j\in\scr M(\tau)$  such that it has a neighbor
$k\notin\scr M(\tau)$, i.e., $x_k(\tau)>x_j(\tau)$ with $k\in\scr N_j(\tau)$ (else the agents of $\cup_{t\geq 0}^\infty \scr M(t)$ would induce a disconnected subgraph, which contradicts the repeatedly jointly strongly connected nature of the neighbor graphs).
From \rep{positive}, it can be seen that $x_j(\tau+1)>x_j(\tau)$.
Using the same arguments, there must exist a finite time $\bar\tau>t$ such that
$\min_{i\in[n]}x_i(\bar\tau)>\min_{i\in[n]}x_i(t)$, which implies that $V(x(\bar\tau))<V(x(t))$.
Therefore, $x_i(t)$ will converge to $1$ for all $i\in[n]$.
\hfill$\qed$

Theorem \ref{mainpositive} implies that system \rep{positive1}
will reach a consensus for any initial condition. Necessary and sufficient conditions for the two extreme opinions
are also given. Specifically, the consensus will be reached at $-1$ if and only if
$x(0)=-\1$, and at $1$ if and only if $x(0)\ll \1$ does not hold.

\begin{remark}
The discrete-time model with stubborn positives has the same limiting behavior
as the continuous-time model considered in \cite{polar}.
We consider the general time-varying case whereas only the time-invariant case
was studied in \cite{polar}.
Moreover, we establish necessary and sufficient conditions for the two extreme opinions, i.e., $-1$ and $1$,
whereas only consensus to $1$ was studied in \cite{polar}.
\hfill$\Box$
\end{remark}

\begin{remark}
We now compare the discrete-time model with stubborn positives \rep{positive} with the original DeGroot model \rep{degroot} for the case in which the neighbor graph does not change over time and is a strongly connected graph $\bbb N$. Thus, the corresponding weight matrix $W$ is also time-invariant. Since $\bbb N$ is strongly connected, $W$ is irreducible. It is well known that in this case, all $x_i(t)$, $i\in[n]$, in the DeGroot model will reach a consensus at value 
$c^\top x(0)$, where $c^\top$ is the unique left eigenvector of $W$ associated with eigenvalue $1$ which satisfies $c^\top \1=1$; moreover, $c\gg \0$. Therefore, as long as there exists at least one agent $i$ for which $x_i(0)<1$, all the agents will not reach a consensus at value $1$. This is a significant difference from the model \rep{positive} in which as long as at least one agent has initial opinion at $1$, all the agents' opinions will converge to $1$.
\hfill$\Box$
\end{remark}

\subsection{Stubborn Neutrals}

Now we consider the case where each individual has susceptibility function $f_i(x_i(t)) = x_i(t)^2$. We call such an individual a stubborn neutral.

\subsubsection*{Motivation}
Observe that for stubborn neutral agents, her susceptibility to being influenced decreases as $x_i \to 0$, and increases as $x_i \to \pm 1$. This means that the closer the individual's opinion is to ``neutral'', i.e. $x_i = 0$,  the more stubborn she becomes. In networks with stubborn neutrals, we consider the neutral opinion as an established, socially normative opinion. For an illustrative example, suppose that the topic was on the level of environmental regulations, e.g. for nuclear power. Then $x_i = 0$ represents individual $i$ favoring the maintaining of the status quo, $x_i = 1$ represents favoring increasing regulation, and $x_i = -1$ represents favoring of decreasing regulation. Some literature showed that pressures existed on individuals in a social network to conform with the group norm \cite{gorden1952interaction_pressure,asch1951group_pressure_effects}, with deviants being punished \cite{thrasher1927gang} or receiving additional pressure to conform \cite{schachter1951deviation}. Merei showed in \cite{merei1949group_leadership} that established traditions heavily influenced the behavior of individuals despite a strong leader attempting to influence change. In the context of our paper, tradition is $x = 0$ and the leader is an individual $i$ with \mbox{$x_i(0) = \pm 1$}, and with $w_{ji}$ large, for any individual $j$ who listens to the leader. Lastly, institutionalization has been linked to the persistence of cultures and resistance to changing the status quo \cite{zucker1977institutuionalization}. In summary, stubborn neutrals may occur in social networks where individuals are reluctant to change from the established norm because of associated risks, or due to institutionalization, or because of pressure to conform.

\subsubsection*{Analysis}
In this case, from \rep{model}, each agent $i$ updates her opinion by setting
\begin{eqnarray}
&&x_i(t+1) \nonumber\\
&=& x_i(t) + x_i(t)^2\sum_{j\in\scr{N}_i(t)} w_{ij}(t)\left(x_j(t)-x_i(t)\right) \nonumber \\
&=& x_i(t) + x_i(t)^2\left(\sum_{j\in\scr{N}_i(t)} w_{ij}(t)x_j(t)-x_i(t)\right) \nonumber \\
&=& x_i(t) - x_i(t)^3 +x_i(t)^2\sum_{j\in\scr{N}_i(t)} w_{ij}(t)x_j(t).
\label{neutral}
\end{eqnarray}
Then, it follows that
\begin{eqnarray}
x(t+1) &=& \left(I - X(t)^2 + X(t)^2 W(t)\right) x(t).  \label{neutral1}
\end{eqnarray}
The following theorem characterizes some limiting behavior of system \rep{neutral1}.

\begin{theorem}
Suppose that Assumption \ref{weight} holds and
that the sequence of neighbor graphs $\bbb N(1),\bbb N(2),\ldots$ is
repeatedly jointly strongly connected.
If $x_i(0) > 0$ for all $i\in[n]$, then all $x_i(t)$, $i\in[n]$, in \rep{neutral}
will reach a consensus exponentially fast at some value in the interval $(0,1]$;
moreover, in this case, the consensus value equals $1$ if, and only if, $x_i(0) = 1$ for all $i\in[n]$.
If $x_i(0) < 0$ for all $i\in[n]$, then all $x_i(t)$, $i\in[n]$, in \rep{neutral}
will reach a consensus exponentially fast at some value in the interval $[-1,0)$;
moreover, in this case, the consensus value equals $-1$ if, and only if, $x_i(0) = -1$ for all $i\in[n]$.
\label{mainneutral}\end{theorem}

{\em Proof:}
From \rep{neutral1}, $x(t+1)=S(x(t),t)x(t)$ where
\eq{S(x(t),t) =  I - X(t)^2 + X(t)^2 W(t).\label{yyy}}
From the proof of Lemma \ref{mono}, $S(x(t),t)$ is a stochastic matrix for all time $t$.

First suppose that $x_i(0) > 0$ for all $i\in[n]$.
There must exist a positive number $\alpha$ such that $x_i(0)\ge \alpha$ for all $i\in[n]$.
It follows from Lemma \ref{mono} that $x_i(t)\ge \alpha$ for all $i\in[n]$ and time $t$.
From \rep{yyy}, we obtain the following two inequalities for $s_{ij}(t)$, the entries of $S(x(t),t)$.
For each diagonal entry,
\begin{eqnarray*}
s_{ii}(t) &=& 1-x_i(t)^2 + x_i(t)^2w_{ii}(t) \\
&\ge& x_i(t)^2w_{ii}(t) \\
&\ge & \alpha^2\beta,
\end{eqnarray*}
where the last inequality makes use of Assumption~\ref{weight} and the fact that we assumed every node in $\bbb N(t)$ has a self-loop, for all $t$.
For each off-diagonal entry,
$$s_{ij}(t)=x_i(t)^2w_{ij}(t).$$
Thus, $s_{ij}(t)$ is nonzero if and only if $w_{ij}(t)$ is nonzero,
which implies that the graph of $S(x(t),t)$ is the same as the graph of $W(t)$, as well as neighbor graph $\bbb N(t)$.
Moreover, it can be seen that when $s_{ij}(t)>0$, it must hold that $s_{ij}(t)\ge \alpha^2 \beta$.
From Proposition \ref{consensus}, all $x_i(t)$, $i\in[n]$, will reach a consensus exponentially fast.
Since $x_i(t)\ge \alpha$ for all $i\in[n]$ and time $t$, the consensus value must lie in $(0,1]$.

Next we show that all $x_i(t)$, $i\in[n]$, in \rep{neutral}
will reach a consensus at value $1$ if and only if $x_i(0) = 1$ for all $i\in[n]$.
Suppose that, to the contrary, there exists at least one $i\in[n]$ such that $0<x_i(0)<1$.
From \rep{neutral}, since $x_j(t)\le 1$ for all $j\in\scr N_i(t)$ implies $\sum_{j\in \scr N_i(t)} w_{ij}(t) x_j(t) \leq 1$,
there holds
\begin{eqnarray*}
x_i(t+1) &\le&  x_i(t)-x_i(t)^3+x_i(t)^2.
\end{eqnarray*}
It can be verified that
$x_i(t)-x_i(t)^3+x_i(t)^2$ increases as $x_i(t)$ increases when $x_i(t)\in(-\frac{1}{3},1)$ and assumes the value $1$ at $x_i(t) = 1$.
Suppose that $0<x_i(t)<1$. Then, it follows that
$x_i(t+1)<1$,
which implies that
if $x_i(0)<1$, then $x_i(t)<1$ for all $t$.
Let $\scr S(t)$ denote the set of agents whose opinions are less than $1$ at time $t$.
From the hypothesis and preceding discussion, $\scr S(t)$ is nonempty for all time $t$.
Let $\bar{\scr  S}(t)=[n]\setminus \scr S(t)$ be the set of agents whose opinions equal $1$ at time $t$.
If $\bar{\scr  S}(t)$ is empty, i.e.,
$x_i(0) < 1$ for all $i\in[n]$, then from Lemma \ref{mono}, the system cannot reach a consensus at $1$.
Suppose that $\bar{\scr  S}(t)$ is nonempty, i.e., there exists at least one agent whose initial opinion is $1$.
Since the sequence of neighbor graphs is repeatedly jointly strongly connected,
there must exist a finite time $\tau$ and an agent $j\in\bar{\scr S}(\tau)$  such that it has a neighbor
$k\in\scr S(\tau)$, i.e., $x_j(\tau)=1$ and $x_k(\tau)<1$ with $k\in\scr N_j(\tau)$.
From \rep{neutral}, and with $i$ replaced by $j$, it can be seen that $x_j(\tau+1)<1$.
Using the same arguments, there exists a finite time $\bar \tau$ such that
$x_i(\bar\tau)<1$ for all $i\in[n]$, which contradicts the hypothesis that
all $x_i(t)$, $i\in[n]$, will reach a consensus at $1$. Therefore,
the consensus at $1$ will be reached if and only if $x_i(0) = 1$ for all $i\in[n]$.

Now suppose that $x_i(0) < 0$ for all $i\in[n]$.
There must exist a positive number $\alpha$ such that $|x_i(0)|\ge \alpha$ (i.e., $x_i(0)\le-\alpha$) for all $i\in[n]$.
It follows from Lemma \ref{mono} that $|x_i(t)|\ge \alpha$ (i.e., $x_i(t)\le-\alpha$) for all $i\in[n]$ and time $t$.
From \rep{yyy}, we obtain the following two inequalities for $s_{ij}(t)$, the entries of $S(x(t),t)$.
For each diagonal entry,
\begin{eqnarray*}
s_{ii}(t) &=& 1-x_i(t)^2 + x_i(t)^2w_{ii}(t) \\
&\ge& x_i(t)^2w_{ii}(t) \\
&\ge & \alpha^2\beta,
\end{eqnarray*}
where the last inequality makes use of Assumption \ref{weight}.
For each off-diagonal entry,
$$s_{ij}(t)=x_i(t)^2w_{ij}(t).$$
Thus, $s_{ij}(t)$ is nonzero if and only if $w_{ij}(t)$ is nonzero,
which implies that the graph of $S(x(t),t)$ is the same as the graph of $W(t)$, as well as neighbor graph $\bbb N(t)$.
Moreover, it can be seen that when $s_{ij}(t)>0$, it must hold that $s_{ij}(t)\ge \alpha^2 \beta$.
From Proposition \ref{consensus}, all $x_i(t)$, $i\in[n]$, will reach a consensus exponentially fast.
Since $x_i(t)\le -\alpha$ for all $i\in[n]$ and time $t$, the consensus value must lie in $[-1,0)$.

Next we show that all $x_i(t)$, $i\in[n]$, in \rep{neutral}
will reach a consensus at value $-1$ if and only if $x_i(0) = -1$ for all $i\in[n]$.
Suppose that, to the contrary, there exists at least one $i\in[n]$ such that $-1<x_i(0)<0$.
From \rep{neutral}, since $x_j(t)\ge -1$ for all $j\in\scr N_i(t)$ implies $\sum_{j\in \scr N_i(t)} w_{ij}(t) x_j(t) \geq -1$,
there holds
\begin{eqnarray*}
x_i(t+1) &\ge&  x_i(t)-x_i(t)^3-x_i(t)^2.
\end{eqnarray*}
It can be verified that
$x_i(t)-x_i(t)^3+x_i(t)^2$ increases as $x_i(t)$ increases when $x_i(t)\in(-1,\frac{1}{3})$.
Suppose that $-1<x_i(t)<0$. Then, it follows that
$x_i(t+1)>-1$,
which implies that
if $x_i(0)>-1$, then $x_i(t)>-1$ for all $t$.
Let $\scr S(t)$ denote the set of agents whose opinions are greater than $-1$ at time $t$.
From the hypothesis and preceding discussion, $\scr S(t)$ is nonempty for all time $t$.
Let $\bar{\scr  S}(t)=[n]\setminus \scr S(t)$ be the set of agents whose opinions equal $-1$ at time $t$.
If $\bar{\scr  S}(t)$ is empty, i.e.,
$x_i(0) > -1$ for all $i\in[n]$, then from Lemma \ref{mono}, the system cannot reach a consensus at $-1$.
Suppose that $\bar{\scr  S}(t)$ is nonempty, i.e., there exists at least one agent whose initial opinion is $-1$.
Since the sequence of neighbor graphs is repeatedly jointly strongly connected,
there must exist a finite time $\tau$ and an agent $j\in\bar{\scr S}(\tau)$  such that it has a neighbor
$k\in\scr S(\tau)$, i.e., $x_j(\tau)=-1$ and $x_k(\tau)>-1$ with $k\in\scr N_j(\tau)$.
From \rep{neutral}, and with $i$ replaced by $j$, it can be seen that $x_j(\tau+1)>-1$.
Using the same arguments, there exists a finite time $\bar \tau$ such that
$x_i(\bar\tau)>-1$ for all $i\in[n]$, which contradicts the hypothesis that
all $x_i(t)$, $i\in[n]$, will reach a consensus at $-1$. Therefore,
the consensus at $-1$ will be reached if and only if $x_i(0) = -1$ for all $i\in[n]$.
\hfill$\qed$

\begin{remark}
It should be noted that Theorem~\ref{mainneutral} does not consider the case where there exist $i,j\in[n]$ such that $x_i(0) > 0$ and $x_j(0) < 0$. It has been shown in \cite{polar} that for this case, the continuous-time model has $\lim_{t\to\infty} x(t) = \0$. However, this is not always the case for the discrete-time model with stubborn neutrals, as we will show shortly,
since in discrete-time, the opinion of an agent $i$ can jump from $x_i(t) > 0$ to $x_i(t+1) < 0$, but this is not possible in the continuous-time case.
Although we will only characterize partial limiting behavior of the discrete-time model for this case (see Theorem~\ref{zero}), we consider the general time-varying graph case whereas only the time-invariant graph case was studied in \cite{polar}. Moreover, we establish necessary and sufficient conditions for the two extreme opinions, i.e., $-1$ and $1$,
which were not provided in \cite{polar}. 
\hfill$\Box$
\end{remark}

The following example shows that 
the discrete-time model with stubborn neutrals \rep{neutral} has different limiting behaviors
from the continuous-time model considered in \cite{polar} (cf. Theorem~5 in \cite{polar}).

\begin{example}
Suppose that there are 4 agents labeled 1 through 4. The neighbor graph is a complete graph and all the weights equal $\frac{1}{4}$. Suppose that the initial opinions are $x_1(0)=1$ and $x_i(0)=-1$ for $i\in\{2,3,4\}$. Thus, 
$$W=
\left[
  \begin{array}{cccc}
    \frac{1}{4} & \frac{1}{4} & \frac{1}{4} & \frac{1}{4} \\
    \frac{1}{4} & \frac{1}{4} & \frac{1}{4} & \frac{1}{4} \\
    \frac{1}{4} & \frac{1}{4} & \frac{1}{4} & \frac{1}{4} \\
    \frac{1}{4} & \frac{1}{4} & \frac{1}{4} & \frac{1}{4} \\
  \end{array}
\right], \;\;\;\;\; 
x(0)=\left[
\begin{array}{c}
1 \\ -1 \\-1 \\-1 \\
\end{array}
\right].
$$
From \rep{neutral}, $x_i(1)=-\frac{1}{2}$ for all $i\in\{1,2,3,4\}$, i.e., all the agents reach a consensus at $-\frac{1}{2}$. 
Similarly, if $x_1(0)=-1$ and $x_i(0)=1$ for $i\in\{2,3,4\}$, with the same weight matrix, all the agents will reach a consensus at $\frac{1}{2}$.
Therefore, in the case when initial opinions contain both positive and negative values, the consensus value can be positive, negative, or zero (as shown in Theorem~\ref{zero}), depending on initial values, as well as the neighbor graph topology.
\hfill$\Box$
\end{example}

\begin{theorem}
Suppose that Assumption~\ref{weight} holds and
that the sequence of neighbor graphs $\bbb N(1),\bbb N(2),\ldots$ is
repeatedly jointly strongly connected.
If $x_i(0) = 0$ for at least one $i\in[n]$, then all $x_i(t)$, $i\in[n]$, in \rep{neutral}
will reach a consensus at value $0$.
\label{zero}\end{theorem}

{\em Proof:}
From \rep{neutral}, if $x_i(t)=0$, then $x_i(t+1)=0$ and thus $x_i(\tau)=0$ for all $\tau\ge t$. 
If all the initial opinions equal $0$, the theorem is clearly true. Suppose therefore that there exists at least one agent whose initial opinion does not equal $0$. 
Consider the
Lyapunov function
$$V(x(t)) = \max_{i\in[n]} |x_i(t)|.$$
From the preceding discussion,
if $x_i(0) = 0$, then $x_i(t)=0$ for all time $t$.
Thus, if $x_i(t)=0$ for all $i\in[n]$ at some time $t$, then
$V(x(t))=0$ and $V(x(t+1))=0$.
Since there exists at least one agent whose opinion initially equals $0$ and thus keeps at $0$, system \rep{neutral1} cannot reach any consensus state (i.e., a state at which all the agents have the same opinion) except for $\0$.
From Lemma~\ref{mono}, $V(x(t+1))\le V(x(t))$. 
Consider a specific time $t$. 
Let $j$ be any agent such that $|x_j(t)|\neq V(x(t))$, which implies that $|x_j(t)|< V(x(t)) \leq 1$.
From the second equality in \rep{neutral}, 
it can be verified that $|x_j(t+1)|<V(x(t))$. 
Let $i$ be any agent such that $|x_i(t)| = V(x(t))$. First consider the case when $x_i(t)>0$, which implies that $x_i(t)=\max_{k\in[n]} x_k(t)$.
Let $\scr M(\tau)$ be the set of agents whose opinions are the largest at time $\tau$,
i.e., $x_j(\tau)=\max_{k\in[n]}x_k(\tau)$ for each $j\in\scr M(\tau)$.
Since $x_i(0) = 0$ for at least one $i\in[n]$,
it follows that $[n]\setminus\scr M(\tau)$ is nonempty unless all the agents reach a consensus at $0$ at time $\tau$.
Since the sequence of neighbor graphs is repeatedly jointly strongly connected,
there must exist a finite time $\tau\ge t$ and an agent $j\in\scr M(\tau)$  such that it has a neighbor
$k\notin\scr M(\tau)$, i.e., $x_k(\tau)<x_j(\tau)$ with $k\in\scr N_j(\tau)$.
From the second equality in \rep{neutral}, with $i$ replaced by $j$, it can be seen that $x_j(\tau+1)<x_j(\tau)$, which implies that $V(x(\tau))<V(x(t))$.
Similarly, the case when $x_i(t)<0$, there must exist a finite time $\bar\tau>t$ such that
$V(x(\bar\tau))<V(x(t))$. 
Therefore, $x_i(t)$ will converge to $0$ for all $i\in[n]$.
\hfill$\qed$

\section{Simulations} \label{sim}
\begin{figure}
\centering
\includegraphics[height=0.85\linewidth,angle=-90]{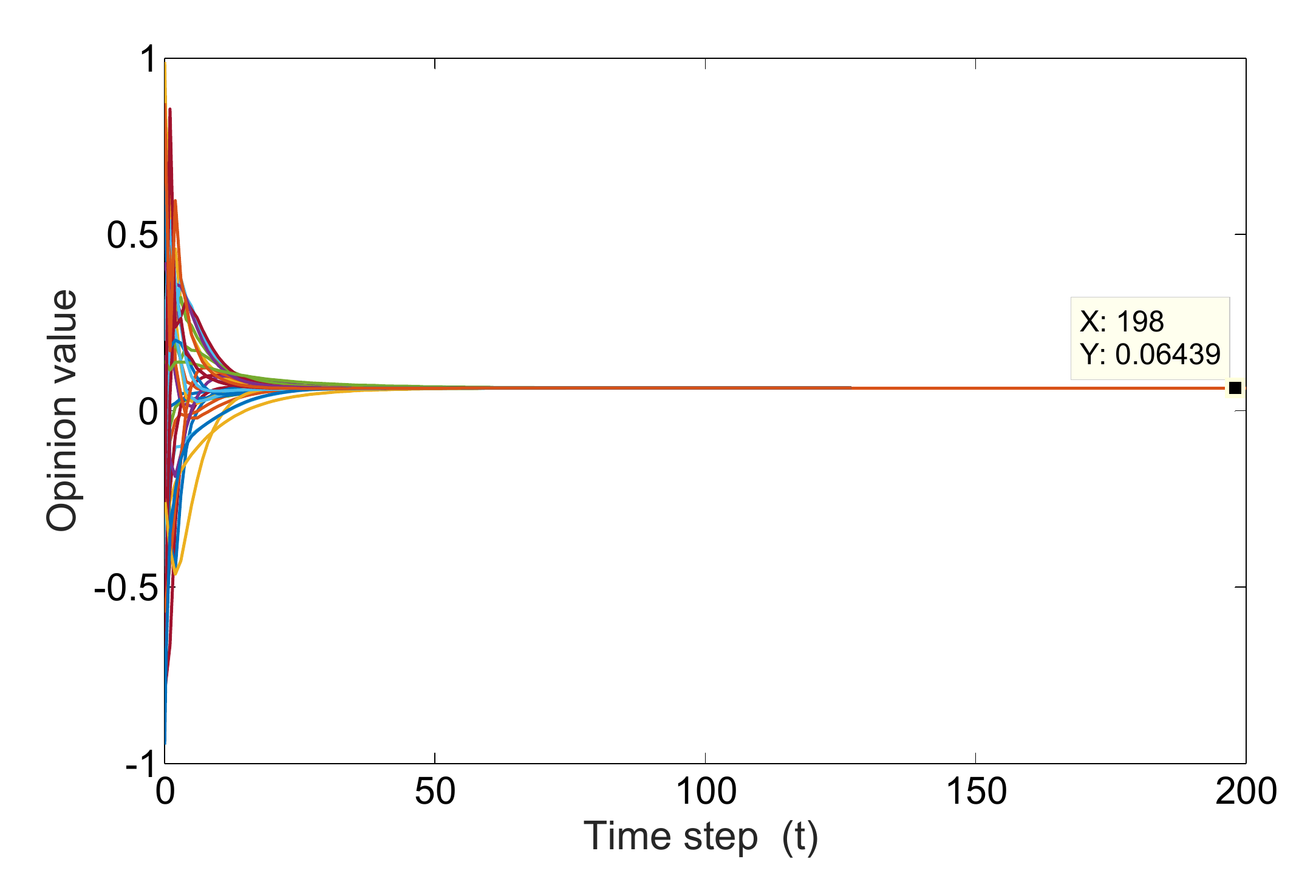}
\caption{Evolution of opinions for individuals fully susceptible (original DeGroot model).}
\label{fig:ACC_DeGroot}
\end{figure}

\begin{figure}
\centering
\includegraphics[height=0.85\linewidth,angle=-90]{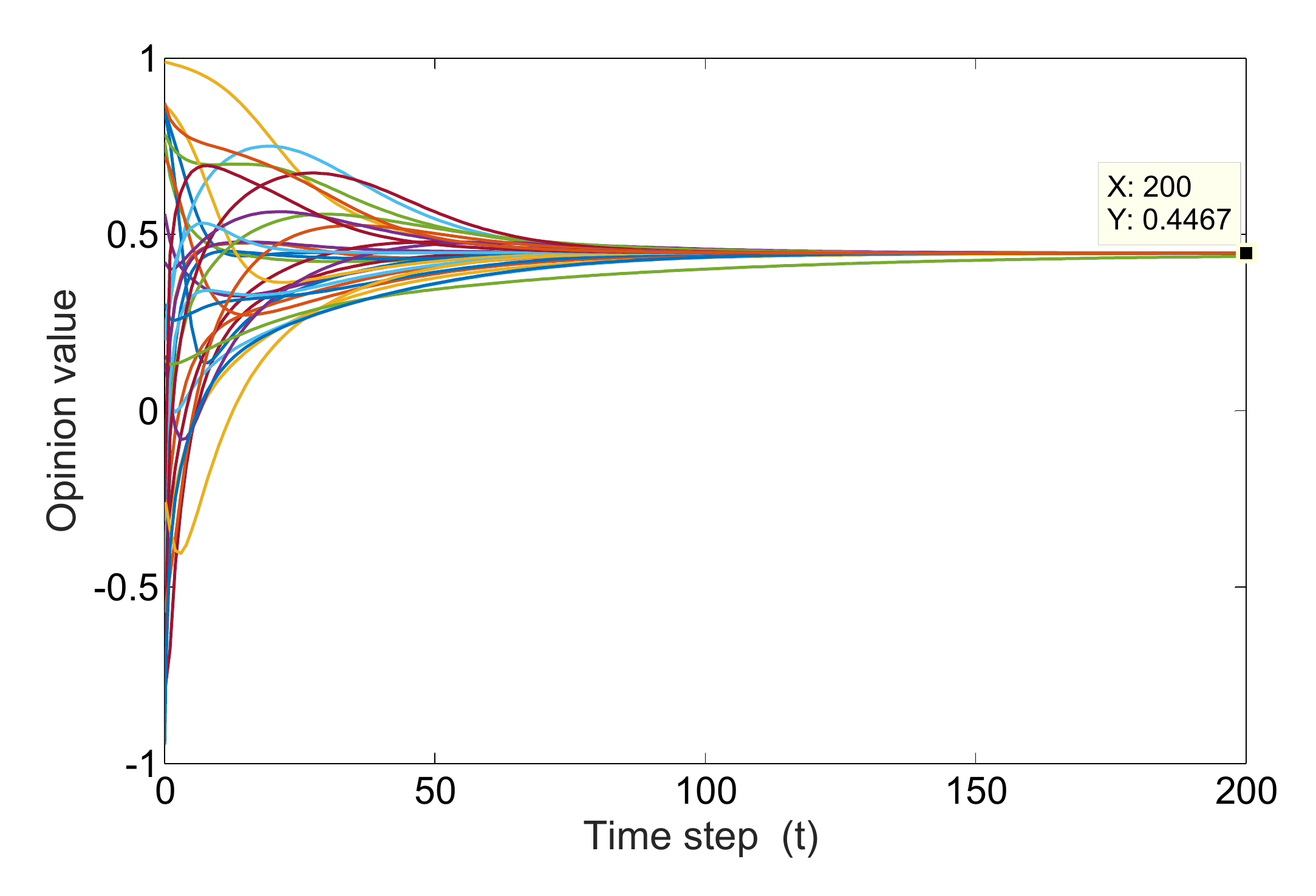}
\caption{Evolution of opinions for stubborn positive individuals ($f_i(x_i(t)) = \frac{1}{2}(1-x_i(t))$).}
\label{fig:ACC_StubbornPositive}
\end{figure}

We now provide a simple simulation example to highlight the effects of the susceptibility function $f_i(x_i(t))$ when compared to the original DeGroot model. We generate a social network with $n=30$ individuals, whose graph is strongly connected, and with randomly selected influence weights $w_{ij}$ which we have assumed are time-invariant for simplicity. We omit showing the $W$ matrix due to spatial limitations. The initial conditions are sampled from a uniform distribution in the interval $(0,1)$. For \emph{the same graph and initial conditions}, we simulated $1)$ the opinion dynamics as modeled by the original DeGroot process, i.e. $f_i(x_i(t)) = 1,\forall\,i\in [n]$, and $2)$ the opinion dynamics where each individual is a stubborn positive, i.e. $f_i(x_i(t)) = \frac{1}{2}(1-x_i(t)),\forall\,i\in[n]$. Due to space limitations, we will not include simulations for stubborn neutral individuals. The original DeGroot dynamics and stubborn positive dynamics are shown in Fig.~\ref{fig:ACC_DeGroot} and Fig.~\ref{fig:ACC_StubbornPositive} respectively.

We clearly see that the final consensus value is different, even though the initial conditions and graph topology are the same. Specifically, the stubborn positive individuals with $x_i(0)$ close to one have low susceptibility, and are reluctant to change their opinions. On the other hand, individuals with $x_i(0)$ close to minus one are significantly more open to influence by others. As a result, the final consensus value is much close in value to one. In other words, while both the DeGroot model and model with stubborn positives reaches a consensus of opinions, the polarity of the stubborn positive individuals results in a more \emph{polarized} final opinion, closer to one end of the opinion spectrum. Moreover, the low susceptibility of the individuals with $x_i(t)$ near one significantly reduces the convergence rate. An interesting future work is to determine quantitatively the effects of different susceptibility functions in shifting the final consensus value, and in altering the convergence rate, when compared to the DeGroot model; the qualitatively effects are obvious from studying the susceptibility functions themselves.


\section{Conclusions} \label{ending}
In this paper, a discrete-time polar opinion dynamics model with susceptibility has been studied. We first proposed a general model and showed it unifies the DeGroot model and the Friedkin-Johnsen model by considering state-dependent susceptibility. We then considered specific susceptibility functions motivated by social examples. Conditions on the time-varying graph topology, and the initial opinion values, are given for the social network to reach a consensus. Necessary and sufficient conditions are given on the initial opinion values for the social network to reach a consensus on an extreme opinion at either end of the opinion interval. For future work, we seek study the case of stubborn extremists, $f_i(t) = 1-x_i(t)^2$, and better understand the behavior of stubborn neutrals when there are initial opinions on either side of $x_i = 0$. In addition, we will aim to generalize stubborn positive, neutral, and extremist functions to so that their values at the key points $x_i = -1, 0, 1$ are fixed but may vary smoothly between these points, and consider mixed individuals in the same network.


\bibliographystyle{IEEEtran}
\bibliography{consensus,social,MYE_ANU}


\end{document}